\documentclass[fleqn,twoside]{article}
\usepackage{espcrc2}

\usepackage{graphicx}
\usepackage[figuresright]{rotating}

\newcommand{\AmS}{{\protect\the\textfont2
  A\kern-.1667em\lower.5ex\hbox{M}\kern-.125emS}}

\title{Measurement of High-Energy Solar Neutrons with SEDA-FIB onboard the ISS}

\author{Y. Muraki\address{Solar-Terrestrial Environment Laboratory, Nagoya University, 
        Nagoya 464-8601, Japan}%
        \thanks{email: muraki@stelab.nagoya-u.ac.jp}, 
  K. Koga\address{Aerospace Research and Development Directorate, JAXA, 
         Tsukuba 305-8505, Japan\\}
  H. Matsumoto$^{b}$,
  O. Okudaira$^{b}$,
  S. Shibata\address{Engineering Science Laboratory, College of Engineering, Chubu University,
            Kasugai 487-0027, Japan\\},
  T. Goka\address{Faculty of Engineering, Tokyo City University, Tokyo 158-8557, Japan\\},
  T. Obara\address{Planetary Plasma and Atmospheric Research Center, Tohoku University, 
      Sendai 980-8578, Japan\\}
  and
  T. Yamamoto\address{Department of Physics, Konan University, Kobe 658-8501, Japan\\}} 
       
\begin{document}

\begin{abstract}
A new type of solar neutron detector (SEDA-FIB) was launched 
on board the Space Shuttle Endeavor on July 16 2009, and began collecting data 
at the International Space Station (ISS) on August 25 2009.   
This paper summarizes four years of observations with the solar 
neutron detector SEDA-FIB (Space Environment Data Acquisition 
using the FIBer detector).  
The solar neutron detector FIB can determine both the energy 
and arrival direction of solar neutrons. 
 In this paper, we first present the angular distribution of neutron 
induced protons obtained in Monte Carlo simulations.   The results are compared 
with the experimental results.
Then we provide the angular distribution of background neutrons
during one full orbit of the ISS (90 minutes).  
Next, the angular distribution of neutrons during the 
flare onset time from 20:02 to 20:10 UT on
March 7 2011 is presented.  It is compared with the distribution
when a solar flare is not occurring.  Observed solar neutrons possibly originated from the M-class solar flares 
that occurred on March 7 (M3.7), June 7 (M2.5), 
September 24 (M3.0) (weak signal) and November
3 (X1.9) of 2011 and January 23 of 2012 (M8.7).   
This marked the first time that neutrons have been observed from M-class solar flares.  
A possible interpretation of the neutron production process will be also provided.
\vspace{1pc}
\end{abstract}

\maketitle

\section{Introduction}
  The first results from 
the solar neutron sensor, SEDA-FIB, onboard the International Space Station were presented at the previous cosmic ray conference in Beijing \cite{bib:koga1}. 
In this paper the current status of the analysis is reported.
  
  The paper consists of the following:  First we give the details of the detector 
and trigger logic. Then 
Monte Carlo results for the angular resolution of neutrons are presented and  
the findings are compared with experimental results.   
We present the angular distribution of background neutrons, which  
are formed by collision of 
galactic cosmic rays with the materials of the space station.  

  Next, we give the angular distribution of neutrons in relation to 
the solar direction  during the flare onset time.  
The results indicate that solar neutrons were successfully detected.   The events are discussed  
together with a possible production mechanism.  
In association with the M-class solar flares during
March 2011 and January 2012, solar neutron events were also observed.  

 Recently, solar activity has once again increased.  In the final part of this paper,
 new results of the analysis of
these flares during April and May of 2013 may be given. 

\section{Details of the Detector}
\subsection{Composition of the sensor}

  The solar neutron detector SEDA-FIB is composed of scintillation bars, each with  
dimensions of 6 mm (width)$\times$3 mm (height)$\times$96 mm (length).   
Each layer in the detector consists of 16 bars, 
so each layer has dimensions of 96 mm$\times$96 mm$\times$3 mm (thickness).  
Such layers are stacked at right angles to each other to form the detector.  A schematic view of the detector is shown in Figure 1.  
  
The signals are read out by using two multi-anode photomultipliers 
(Hamamatsu R4140-20MOD) with 256 read-out channels. 
(Initially, we were concerned about discharge at the corner of the
photomultipliers, but this has not had any serious effects on observations.)
The detector records not only the coordinates of the track 
of the detected particle but also the energy deposited in 
each bar.   With this information, we could distinguish 
neutron-induced protons from photon-induced electrons.    

When neutron-induced protons stop in a scintillator bar, 
they deposit the bulk of their energy within in.   
From the Bragg peak, we can 
determine the direction of the incoming neutrons.  
An actual example of an event is presented in Figure 2.
The signal threshold of the  ADC channel is set above $\sim$11 
and the peak value for a minimum ionizing particle (MIP) is set at around $\sim$22.  
The value varies between channels.
The red color in Figure 2 corresponds to a ADC channel value of 64(G-max) and
the blue color corresponds to 11 (G-min).  This means that an ADC channel value of 66 
corresponds to six times the value of the MIP. 
The sensor has a cubic shape and its volume is approximately $1,000 cm^{3}$.   
The six faces of the cubic detector are covered by 
six pieces of scintillator plate with size
of 10 cm$\times$10 cm$\times$ 1 cm (thickness).   
These scintillator plates are used as an anti-counter to remove the effect of
other charged particles on the results.  They work very well and protons and 
electrons are rejected very efficiently.  The signals due to neutron-induced protons
are sent to the two photomultipliers via 512 optical fibers, each 1 mm in
diameter.  The two photomultipliers are orientated in the X-Z 
and Y-Z planes of the sensor respectively. 
Here the X-direction is defined as the direction to the Earth$'$s center and the
Y-direction is opposite to the direction 
of motion of the ISS, which is perpendicular to the X-direction 
as the ISS is orbiting the Earth.     Small holes (2$\times$256 holes) 
run through the two scintillator plates to accommodate the 
optical fibers.  
Further details of the detector have been published elsewhere \cite{bib:imaida},
\cite{bib:koga2},\cite{bib:muraki}. 
 
\subsection{Trigger of signals}

The trigger signals are produced by the dynode signals of each photomultiplier.  
When the sum of the dynode signals exceeds 35 MeV, a trigger signal is created 
and the amount of energy deposited in all the scintillator bars is recorded.   
When the trigger rate exceeds 16 counts/sec, only the
 coordinates of the tracks are recorded;  
however, this function actually works when the trigger rate
exceeds about 2 counts/sec.    If the count rate exceeds 64 counts/sec, 
only the deposited energy in each dynode is recorded. This may happen 
when the sensor passes through the South Atlantic Anomaly region or during a period of high neutron flux.  
The typical trigger rate over the Equator is 0.07 counts/sec, while over the North or 
South polar regions it is 0.4 counts/sec. The average trigger rate is 
about 0.22 counts/sec.   The trigger rate sometimes exceeds over 20 counts/sec 
above the SAA region.

The correspondence between the sum of the ADC values 
and the dynode signal is quite good.
The ADC signals are analyzed after removing the background 
pedestal value from each channel.  
The pedestal and the peak values 
for the MIPs of all 512 channels 
are estimated from the results of proton beam experiments at Riken, Japan and
 cosmic ray muon data before launching.

\subsection{Pointing Ability of the Sensor to Determine the Direction of Incoming Neutrons }

   To distinguish solar neutrons from background radiation, 
the pointing ability of the sensor is very important.    
If observations of solar neutrons are made in the background free space 
between the Sun and the Earth, it is not necessary to take this into account;   
however, in our experiment on the ISS, the process becomes very important.  
We have estimated the power by both actual experiments and by MC calculations. 
The experiments were made in 1999 and 2000 
using the RCNP neutron beam at Osaka University \cite{bib:sohno}.  
In this paper we present the angular distribution as predicted by  
the MC simulation based on the Geant-4 program. 
The results are compared with the experimental results.

Figure 3 shows MC results for the angular distribution of protons 
induced by vertically incident
neutrons from above the sensor with 
an energy of 80 MeV.  
The black circles correspond to events induced by neutrons 
inside the cubic sensor (the dimensions of which are 10 cm$\times$10 cm$\times$10 cm) 
without using the anti-counter, while the white squares represent the signals 
recorded by the sensor with the use of the anti-counter,
i.e., neutron-induced protons are involved inside the sensor.   
To produce these results,  one million neutron events were generated.   
The simulation was made for different incident
energies, from 60 to 200 MeV; however, the general character of the results 
did not change with incident energy.

The MC results were next compared with the experimental results.
The experimental results shown in Figure 4 are for an
incident energy of En=80-100 MeV.  Through comparison of Figure 3 and  4, 
 it is apparent that the MC results reproduce the experimental results quite well.

The MC simulation predicted peaks of scattered protons at around 20 degrees 
from the direction of the incident neutrons.
The distribution spreads from 0 to 50 degrees.
This implies that solar neutrons and background neutrons can be distinguished if we only consider neutron induced protons within a cone of 30 degrees
around the solar direction. 
This reduces the effect of background neutrons by 
approximately 16 times,  if we assume the background 
neutrons come equally from all directions.

The detection efficiency for neutrons with deposited energy over of 35 MeV 
is predicted by the MC calculation
to be 0.02 at 60 MeV, 0.023 at 80 MeV, 0.02 at 100 MeV, 
and 0.015 at 140 MeV respectively.  
The experimental results give 
values of 0.02, 0.018, 0.014 and 0.003 respectively.  
Thus, there is good agreement between 
the MC numerical and experimental results.

\section{Angular Distribution of Background Neutrons}

We have analyzed the angular distribution of background neutrons.  
The ISS completes an orbit of the Earth every 90 minutes.  
Therefore we considered the angular distribution of background neutrons about  
the Z-axis over this time period.  
  Figure 5 shows the distribution on 
the {$\theta$-$\phi$}plane.
Figure 6 shows the energy distribution of background-neutron-induced protons 
measured by the {$\it range method$}. 
Figure 7 shows the same plot as Figure 5, but for the short period
21:32-21:40 UT, which is 90 minutes after
the flare onset time of 20:02-20:10 UT.   This data is used to estimate
the background during the flare onset time.

\section{Solar Flares observed on March 7, 2011} 

  On March 7 2011, a middle class solar flare was observed.
According to GOES satellite data,
the intensity of the flare was M3.7; it started at 19:43 UT
and reached a maximum at 20:12 UT.  Since launching, we have tried to observe every solar
flare with an intensity above M2, as recorded by GOES observations.  
  This was the first solar flare for which our sensor detected signals associated with solar neutrons. The signal was not extremely strong
but the statistical significance was 8{$\sigma$}.   

In the previous conference in Beijing, we were surprised to learn that
the Fermi-LAT satellite observed long lasting GeV gamma-rays
as a result of this flare \cite{bib:tanaka}.   
It was unclear until now 
if solar neutrons were observed during the very intensive solar 
flares.  The Solar Dynamical Observatory data indicated an energetic
coronal mass ejection (CME) in association with this flare \cite{bib:cheng}.   
Therefore we think that protons in the CME were accelerated and back scattered, causing some protons 
to hit the solar surface, producing neutrons and
gamma-rays.  More detail of the production mechanism has been
provided in another paper \cite{bib:muraki}. 

   In Figure 8, we present the signals detected by the SEDA-FIB.
To estimate the background involved in the solar cone, it is
useful to use the data after 90 minutes presented in Figure 7.
The expected arrival direction of neutron-induced protons is shown in Figure 8
by the red line.   At that time, the direction of the Sun
was 57 degrees on the
Y-Z plane, which hardly changed during the time of measurement.  
We have plotted the solar position from 19:49 to 20:02 UT by
the open brown squares in Figure 8.   Solar neutron with 80 MeV
need 13 minutes more flight time than the light.

The same selection criteria was applied to the data sets 
between 20:02-20:10 UT (flare time) and 21:32-2140 UT (
off-flare time as presented in Figure 7).
Then, we can say that solar neutrons have been
observed with a statistical significance of 8 {$\sigma$}.



%
\begin{figure}[t]
 \centering
\includegraphics[width=0.33\textwidth]{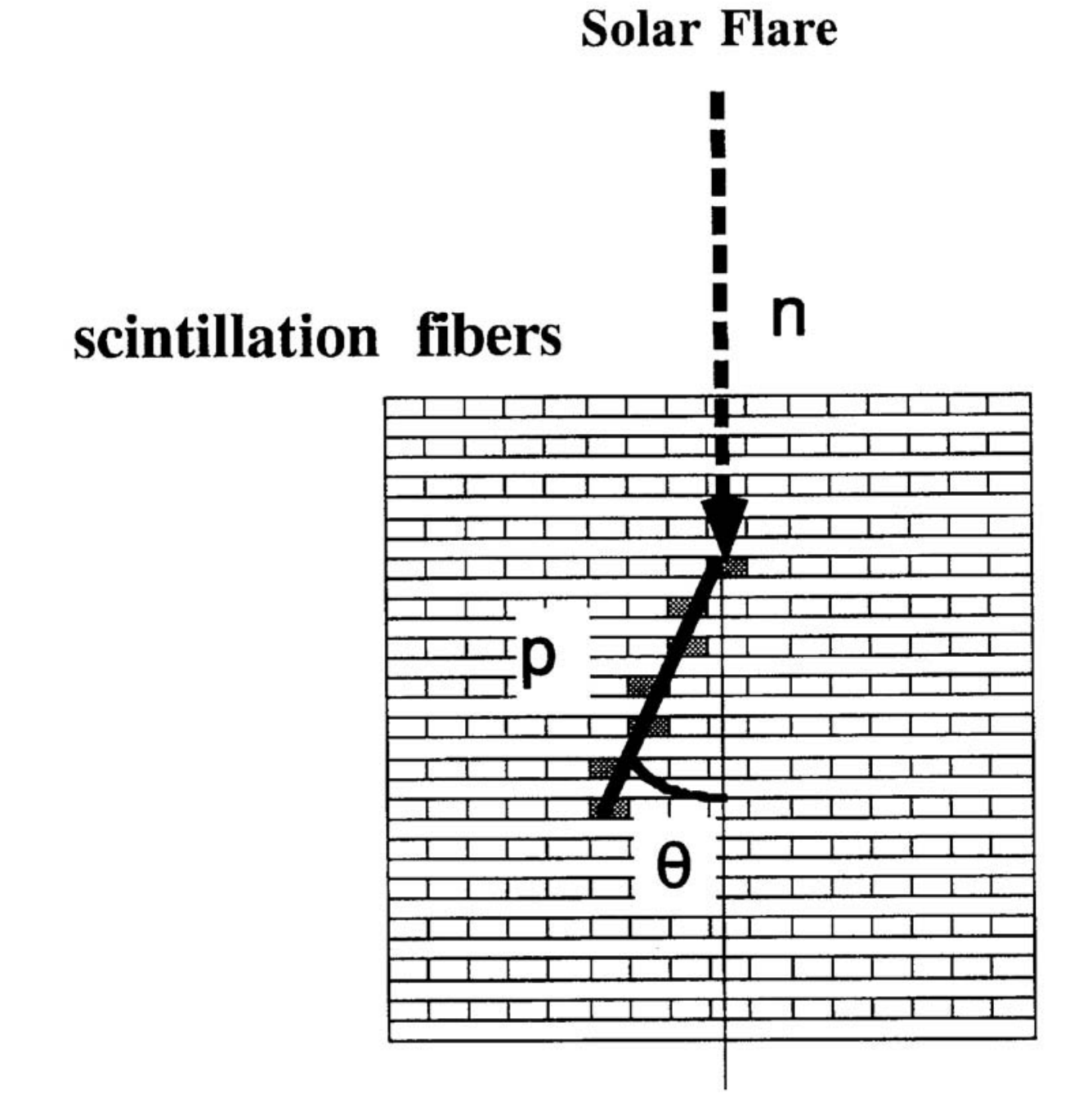}
\caption{The schematic view of the sensor of SEDA-FIB.
  The FIB detector is composed of 16 layers of X-segmented and 16 layers of Y-segmented
  scintillation arrays.   The thickness of one layer is 3 mm. 
 The sensor can identify the incoming direction
  of neutrons using this segmented structure. }
 \label{simp_fig}
\end{figure}
 
\begin{figure}[t]
 \centering
  \includegraphics[width=0.43\textwidth]{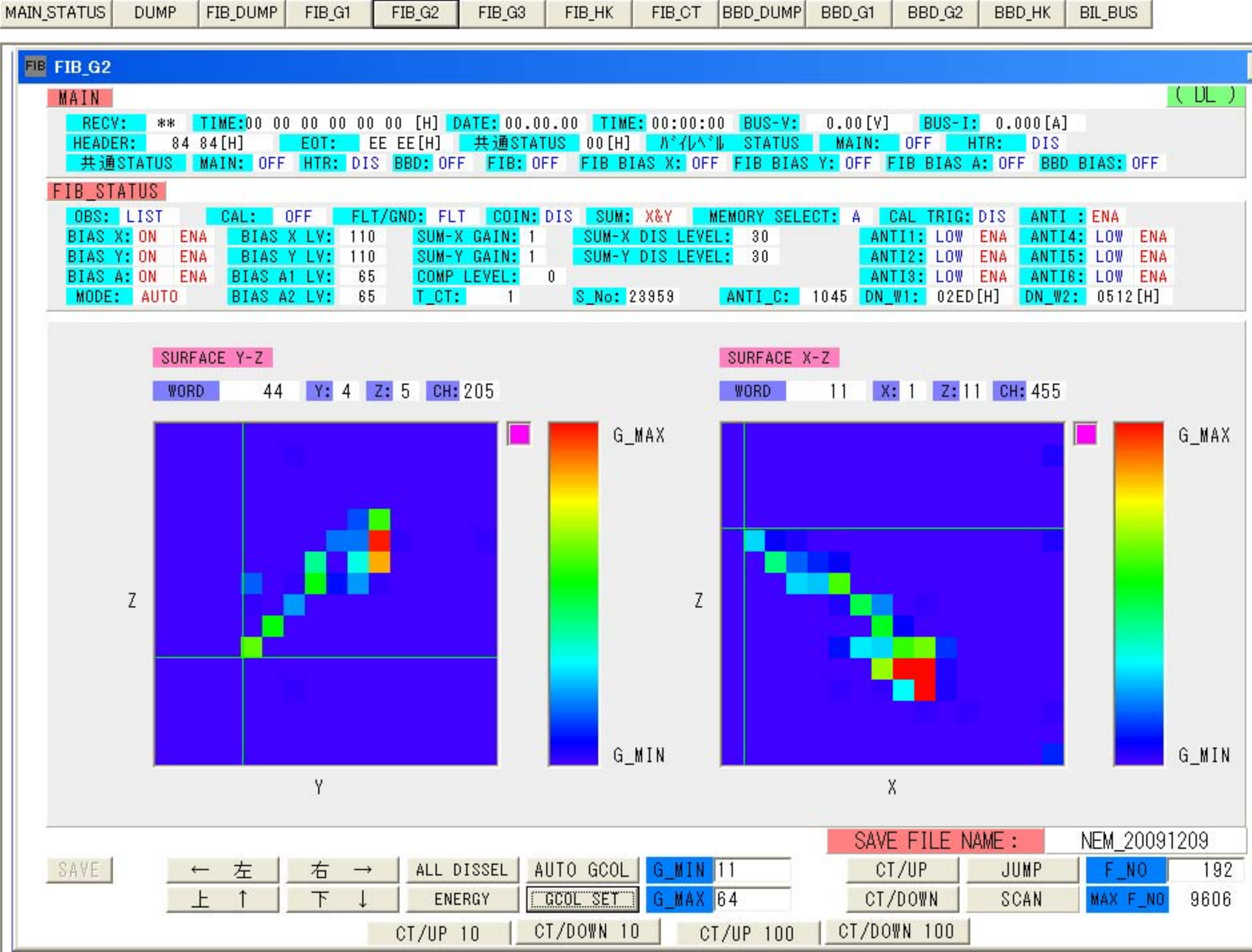}
 \caption{Typical example of neutron detected by SEDA-FIB.  
 A Bragg peak can be recognized.
   The color corresponds to the energy loss of protons in each scintillation bar.}
  \label{simp_fig}
\end{figure}

 \begin{figure}[t]
  \centering
  \includegraphics[width=0.45\textwidth]{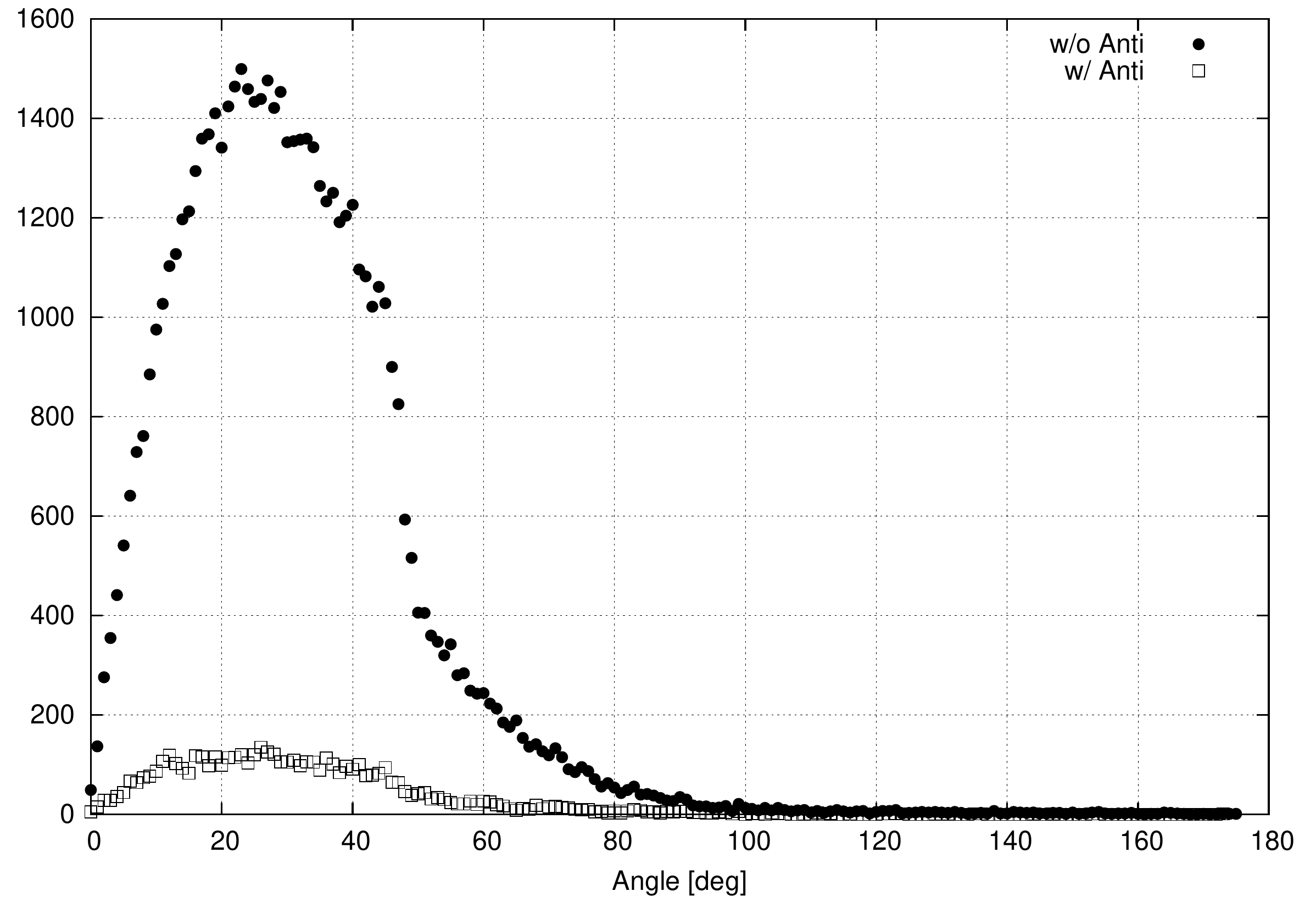}
  \caption{MC results for the expected angular distribution of protons
  induced by incoming neutrons inside the sensor 
  with an incident energy of En=80 MeV.
  The black circles and white squares correspond to events
  without the anti-counter and with the anti-counter, respectively.}
  \label{simp_fig}
 \end{figure}

\begin{figure}[t]
 \centering
  \includegraphics[width=0.45\textwidth]{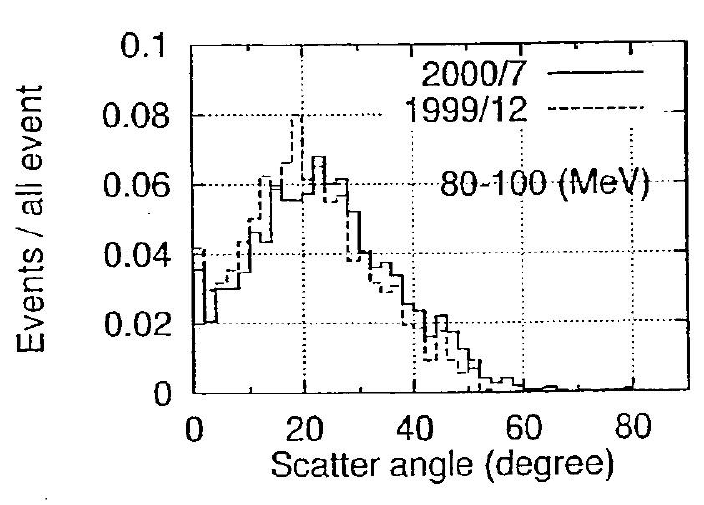}
  \caption{Experimental results for angular distribution of protons
   induced by incoming neutrons with the incident energy of 80-100 MeV.
    The vertical axis represents the number of events
     normalized by the total event, while the horizontal axis
     presents the scattering angle in the unit of degrees. }
 \label{simp_fig}
 \end{figure}

\begin{figure}[t]
 \centering
  \includegraphics[width=0.42\textwidth]{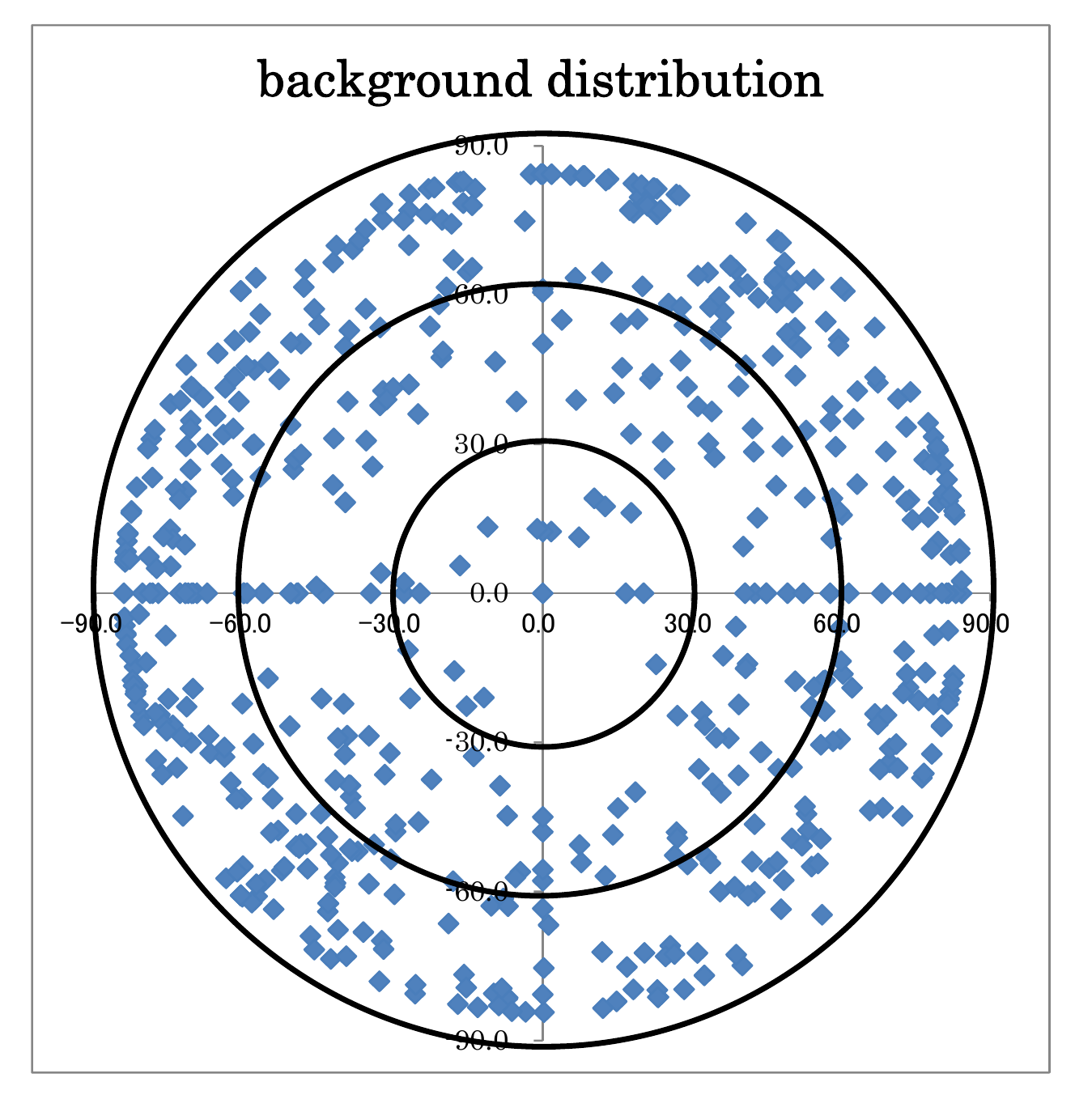}
  \caption{Angular distribution of background-neutron-induced protons
  plotted on the {$\theta$}-{$\phi$ }plane. The data was taken during
  one orbit (90 minutes) of the ISS.  A uniform distribution can be seen.  }
  \label{simp_fig}
 \end{figure}
 
 \begin{figure}[t]
  \centering
  \includegraphics[width=0.40\textwidth]{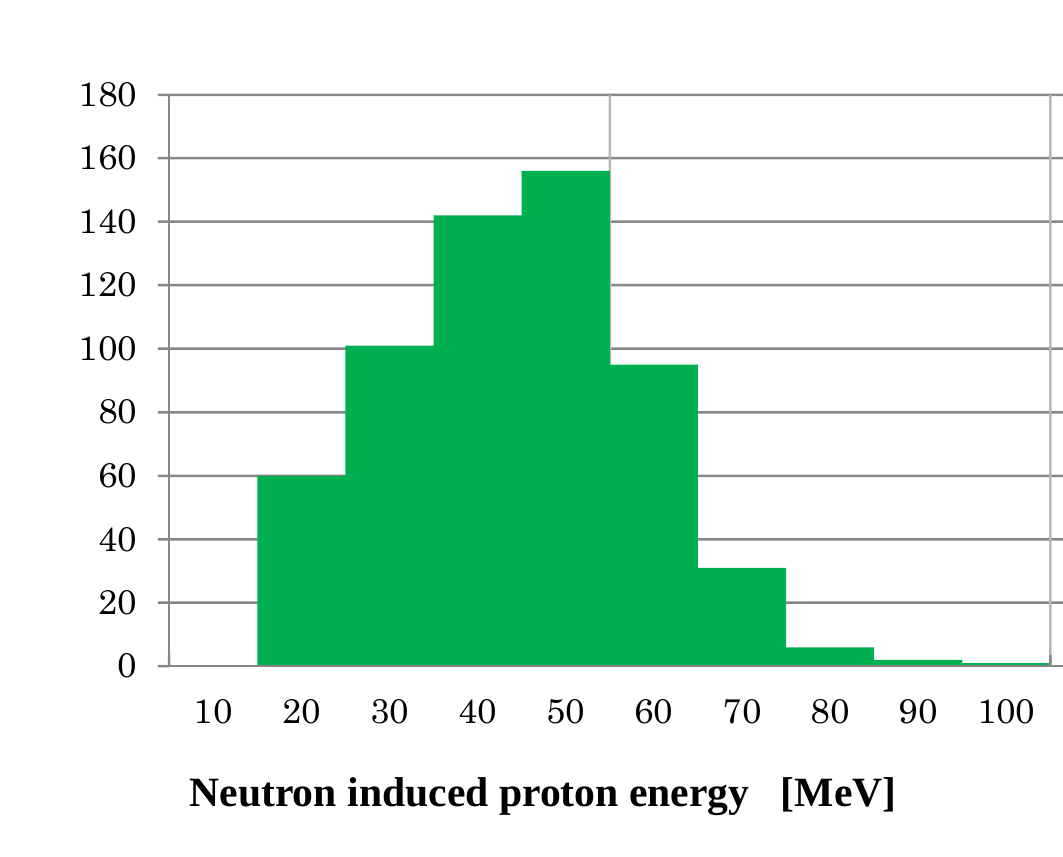}
  \caption{Energy distribution of above 90 minutes background-neutron-induced protons, 
    based on the measured range. The horizontal axis represents 
   the proton kinetic energy in MeV, while the vertical axis
   indicates the number of events.   }
 \label{simp_fig}
 \end{figure}
 
 \begin{figure}[t]
 \centering
  \includegraphics[width=0.42\textwidth]{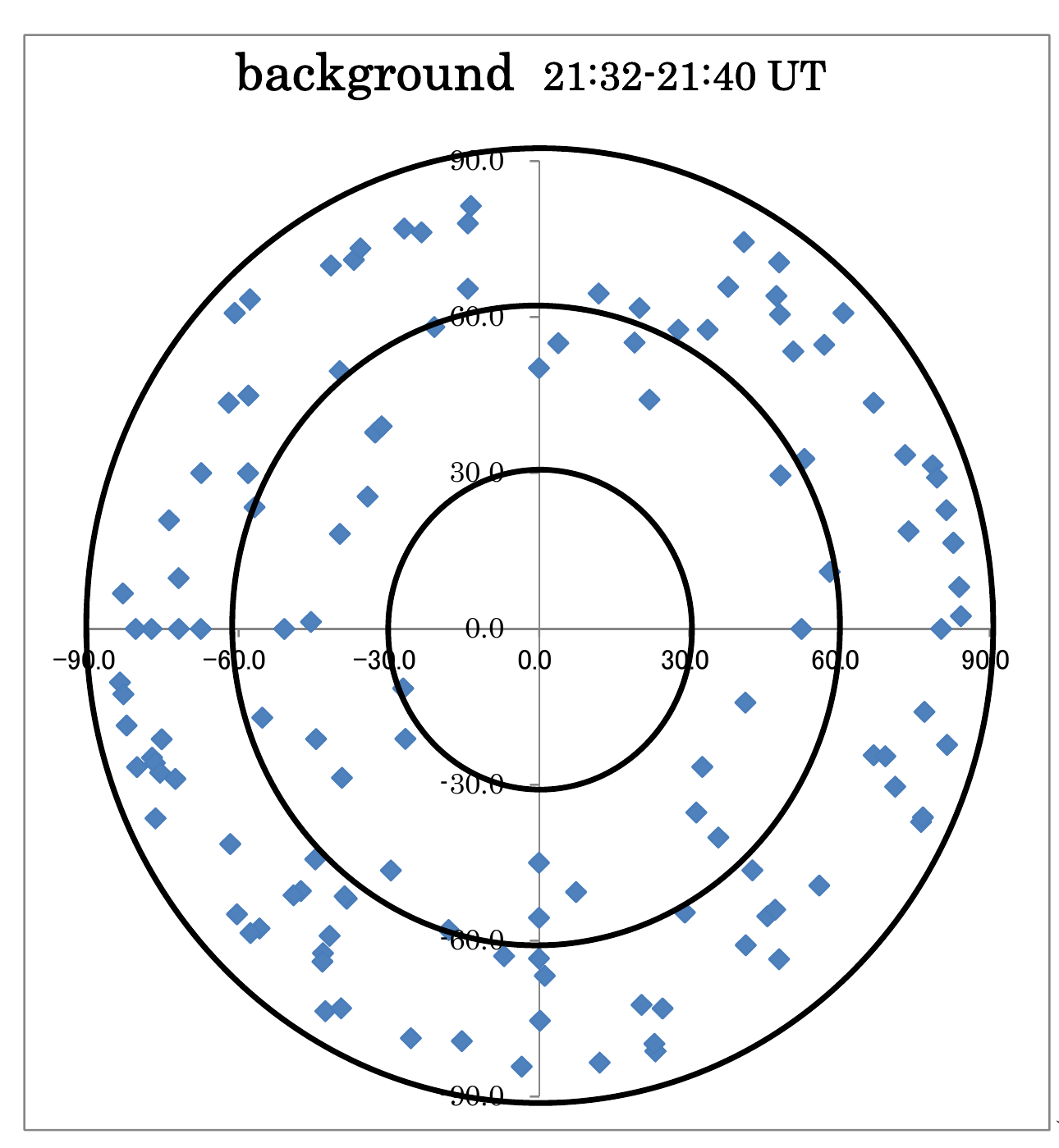}
  \caption{Angular distribution of background-neutron-induced protons
  plotted on the {$\theta$}-{$\phi$ }plane. The data was taken 90 minutes
  after the flare onset time of 21:02-21:40 UT.  }
  \label{simp_fig}
 \end{figure}
 
 \begin{figure}[t]
 \centering
  \includegraphics[width=0.42\textwidth]{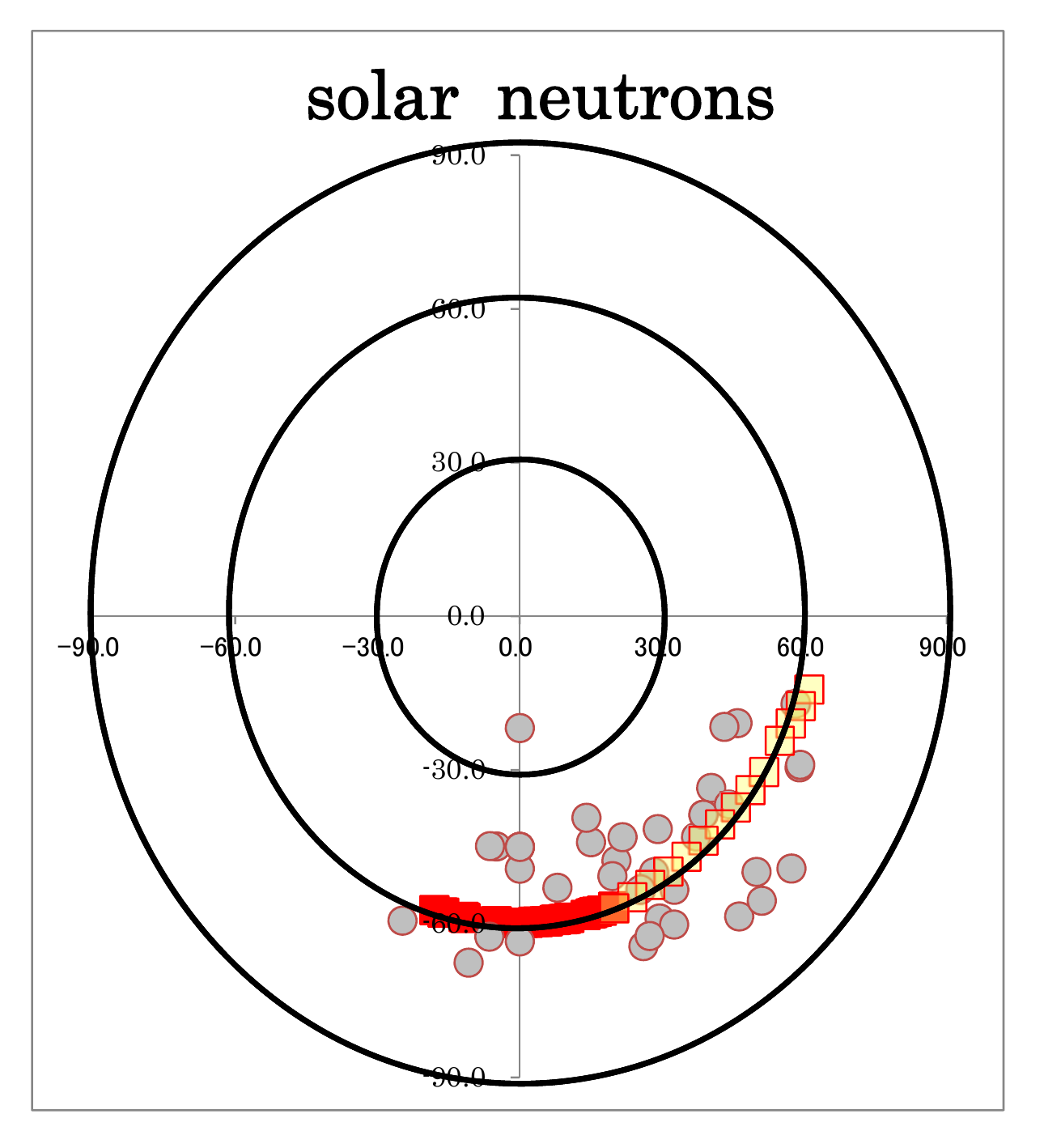}
  \caption{Angular distribution of neutron-induced protons
  plotted on the {$\theta$}-{$\phi$ }plane. The data was taken during
  the flare onset time 20:02 to 20:10 UT on March 7 2011.  
  The the red line represents the direction of the Sun from 20:02
  to 20:10 UT, while open brawn squares present the solar position
  from 19:49 to 20:02 UT when neutrons departed from the Sun.  }
  \label{simp_fig}
 \end{figure}

\section{Conclusions}
  Solar neutrons were detected by the SEDA-FIB onboard the ISS.
This may be the first time solar neutrons from M-class solar flares have been detected.   
From the data observed by the
Fermi-LAT satellite, it is probable that these neutrons were
produced by the back scattered protons accelerated in a CME
by the shock acceleration mechanism.   They are not produced by
the impulsive phase associated with the
largest class solar flares as previously observed.

\vspace*{0.5cm}
\footnotesize{{\bf Acknowledgment:}{ The authors acknowledge the members
of the Tsukuba operation center of Kibo for
taking the SEDA-FIB data every day.   The authors also thank Dr. M. Fujii 
of FAM science for providing software.  }

\end{document}